\documentclass[prd,showpacs,showkeys,floatfix,nofootinbib,eqsecnum,
               preprint,12pt,tightenlines,fleqn]{revtex4-1}

\usepackage{amssymb,epsf}
\usepackage{latexsym,amssymb}
\usepackage{amsmath,mathrsfs}
\usepackage{graphics,epsfig}
\usepackage[english]{babel}

\renewcommand{\baselinestretch}{1}  
\renewcommand{\baselinestretch}{1.125}
\renewcommand{\baselinestretch}{1.08}

\newcommand{\beq}{\begin{equation}}
\newcommand{\eeq}{\end{equation}}
\newcommand{\beqa}{\begin{eqnarray}}
\newcommand{\eeqa}{\end{eqnarray}}
\newcommand{\bsubeqs}{\begin{subequations}}
\newcommand{\esubeqs}{\end{subequations}}

\begin{document}
%
%
\noindent  
Phys. Rev. D 97, 124047 (2018)    
\hfill    
arXiv:1803.09736  
\newline\vspace*{4mm}
%

\title{Antigravity from a spacetime defect \vspace*{4mm}}

\author{F.R. Klinkhamer}
\email{frans.klinkhamer@kit.edu}
\affiliation{\mbox{Institute for Theoretical Physics,
Karlsruhe Institute of Technology (KIT),}
76128 Karlsruhe, Germany\\}
\author{J.M. Queiruga}
\email{jose.queiruga@kit.edu}
\affiliation{\mbox{Institute for Theoretical Physics,
Karlsruhe Institute of Technology (KIT),}
76128 Karlsruhe, Germany\\}
\affiliation{\mbox{Institute for Nuclear Physics,
Karlsruhe Institute of Technology  (KIT),}
\mbox{Hermann-von-Helmholtz-Platz 1,
76344 Eggenstein-Leopoldshafen, Germany}\\}

\begin{abstract}
\noindent
\vspace*{-4mm}\newline
We argue that there may exist
spacetime defects embedded in Minkowski spacetime,
which have negative active gravitational mass.
One such spacetime defect then repels a test particle,
corresponding to what may be called ``antigravity.''
\end{abstract}


\maketitle

\section{Introduction}\label{sec:Introduction}

It is possible that a ``quantum-spacetime'' phase has given
rise to the classical spacetime  of the present Universe.
If that is the case, then it is also possible that there
have appeared small imperfections in the emerging classical
spacetime. These remnant imperfections may be called
\mbox{``spacetime defects,''} by analogy with defects
in an atomic crystal.

We have studied a soliton-type spacetime defect
which has nontrivial topology for both
the spacetime manifold and the matter-field configuration.
Specifically, we have constructed a new type of
Skyrmion classical solution~\cite{Klinkhamer2014-prd},
for which the nontrivial topology of the $SO(3)$ matter field matches
the nontrivial topology of
space [essentially $\mathbb{R}P^3 \sim SO(3)$].

Even before the final soliton was constructed, there
were hints that the soliton could have an unusual
asymptotic property, namely,
a negative active gravitational mass~\cite{Klinkhamer2013}.
This was confirmed by a detailed numerical analysis~\cite{Guenther2017}.
The goal, now, is to understand the origin of this
negative gravitational mass.

\section{Theory and Ans\"{a}tze}
\label{sec:Theory-etc}

The details of the spacetime manifold and matter fields of the
Skyrmion spacetime defect solution have been given in Sec.~II of
Ref.~\cite{Klinkhamer2014-prd}, which also contains
further references. Here, we recall the main
results, mainly in order to establish our notation.

The four-dimensional spacetime manifold has
the topology
\bsubeqs\label{eq:M4-M3-topology}
\beqa\label{eq:M4}
M_4 &=& \mathbb{R} \times M_3\,,
\\[2mm]
\label{eq:M3}
M_3 &=&
\mathbb{R}P^3 - p_\infty\,,
\eeqa
\esubeqs
where $p_\infty$ corresponds to the ``point at spatial infinity.''

A particular covering of the manifold $M_3$
uses three charts of coordinates, labeled by $n=1,\,2,\,3$.
The explicit coordinate charts are discussed in
Sec.~2 and App.~A of Ref.~\cite{Klinkhamer2014-mpla}.
These three sets of coordinates $(X_n,\, Y_n,\,Z_n)$
are invertible and infinitely-differentiable functions of each
other in their respective overlap regions
(details and further results can be found in
Chap.~5 of Ref.~\cite{Schwarz2010}
and a brief summary appears in
Sec.~2.1.2 of Ref.~\cite{Guenther2017}).
At this moment, we will focus on the $n=2$ chart,
the other charts being similar.
Specifically, the chart-2 coordinates have the following ranges:
\beqa\label{eq:X2Y2Z2-ranges}
X_{2} \in (0,\,\pi)\,,\quad
Y_{2} \in (-\infty,\,\infty)\,,\quad
Z_{2} \in (0,\,\pi)\,,
\eeqa
where $Y_2=0$ gives the position of the defect surface
(an $\mathbb{R}P^2$ submanifold). 
Henceforth, we will drop the suffix $2$.

The action reads ($c=\hbar=1$) 
\bsubeqs\label{eq:action-omegamu}
\beqa\label{eq:action}
\hspace*{-8mm}
 S &=&\int_{M_4} d^4X\,\sqrt{-g}\,
\Bigg[
\frac{1}{16\pi G_{N}}\:R
+\frac{f^2}{4}\:\text{tr}\Big(\omega_\mu\,\omega^\mu\Big)
+\frac{1}{16 e^2}\: \text{tr}\Big(\left[\omega_\mu,\,\omega_\nu\right]
\left[\omega^\mu,\,\omega^\nu\right]\Big)\Bigg],
\\[2mm]
\label{eq:omegamu}
\hspace*{-8mm}
\omega_\mu &\equiv& \Omega^{-1}\,\partial_\mu\,\Omega\,,
\eeqa
\esubeqs
in terms of the Ricci curvature scalar $R$ and
the scalar field $\Omega(X)\in SO(3)$, together with the
definition $g\equiv \det g_{\mu\nu}$.
In addition to having Newton's gravitational coupling constant $G_{N}$,
there is an energy scale $f$ from the kinetic matter term in \eqref{eq:action}.
Furthermore, the quartic Skyrme term
comes with a dimensionless coupling constant $1/e^2$.

The spherically symmetric \textit{Ansatz} 
for the metric over the chart-2 domain is given by~\cite{Klinkhamer2014-prd}
\bsubeqs\label{eq:metric-Ansatz-W-definition}
\beqa\label{eq:metric-Ansatz}
\hspace*{-4mm}
ds^2 &=&
 -\big[\mu(W)\big]^2\, dT^2
 +\big(1-b^2/W\big)\,\big[\sigma(W)\big]^2\,(d Y)^2
+W \Big[(d Z)^2+\sin^2 Z\, (d X)^2 \Big]\,,
\\[2mm]
\hspace*{-4mm}
\label{eq:W-definition}
W &\equiv& b^2+Y^2\,.
\eeqa
\esubeqs
Remark that, compared to Eq.~(8) of Ref.~\cite{Klinkhamer2014-prd},
we have dropped the suffix $2$ on the spatial coordinates
in \eqref{eq:metric-Ansatz-W-definition}
and have removed the tildes on 
the two finite metric functions, $\mu(W)$ and $\sigma(W)$. 
The metric from the \textit{Ansatz}
\eqref{eq:metric-Ansatz-W-definition}
is degenerate at $W=b^2$ (or $Y=0$), with a vanishing
determinant of the metric at the defect surface.
This particular \textit{Ansatz} avoids running into curvature
singularities; see Ref.~\cite{KlinkhamerSorba2014} for a discussion of
manifolds with the same topology \eqref{eq:M4-M3-topology} but
inequivalent differential structures.
For further remarks on degenerate metrics in general relativity,
see Sec.~\ref{sec:Discussion}.

The Skyrmion-type \textit{Ansatz} for the $SO(3)$ scalar field
is given by~\cite{Klinkhamer2014-prd}
\bsubeqs\label{eq:hedgehog-Ansatz}
\begin{eqnarray}\label{eq:hedgehog-Ansatz-Omega}
\Omega(X) &=&
\cos\big[F(W)\big]\;\openone_{3}
-\sin\big[F(W)\big]\;
\widehat{x}\cdot \vec{S}
+\big(1-\cos\big[F(W)\big]\big)\;
\widehat{x} \otimes \widehat{x}\,,
\\[2mm]\label{eq:hedgehog-Ansatz-bcs}
F(b^2) &=& \pi\,,\quad F(\infty) = 0\,,
\\[2mm]
S_1 &\equiv&  \left(
                \begin{array}{ccc}
                  0     & 0  &   0 \\
                  0     & 0  &   1 \\
                  \;0\; & -1 & \;0\; \\
                \end{array}
              \right)\,,
\quad
S_2 \equiv  \left(
                \begin{array}{ccc}
                  \;0\; & \;0\; & -1 \\
                    0   &   0   & 0 \\
                    1   &   0   & 0 \\
                \end{array}
              \right)\,,
\quad
S_3 \equiv  \left(
                \begin{array}{ccc}
                  0  &   1   &   0 \\
                  -1 & \;0\; & \;0\; \\
                  0  &   0   &   0 \\
                \end{array}
              \right)\,,
\end{eqnarray}
\esubeqs
with the unit 3-vector $\widehat{x} \equiv \vec{x}/|\vec{x}|$
from the Cartesian coordinates $\vec{x}$ defined in terms
of the coordinates $X$, $Y$, and $Z$
(see App.~A of Ref.~\cite{Klinkhamer2014-mpla}).
Remark that, compared to Eq.~(9) of Ref.~\cite{Klinkhamer2014-prd},
we have removed the tilde on the \textit{Ansatz}
function $F(W)$ in \eqref{eq:hedgehog-Ansatz}.
The boundary conditions \eqref{eq:hedgehog-Ansatz-bcs}
make for unit winding number of the compactified map
$\overline{\Omega(X)}$.
Further details and discussion can be found in
Ref.~\cite{Klinkhamer2014-prd}.

There are two dimensionless model parameters in
the theory \eqref{eq:action-omegamu},
\bsubeqs\label{eq:dimensionless-model-param}
\begin{eqnarray}\label{eq:dimensionless-eta}
\widetilde{\eta}&\equiv& 8\pi\, G_{N}\, f^2 \geq 0\,,
\\[2mm]
e &\in&  (0,\,\infty)\,.
\end{eqnarray}
\esubeqs
Regarding the inequality of \eqref{eq:dimensionless-eta},
we assume $f^2>0$ and allow for $G_{N} \geq 0$.
Numerical calculations can use the following dimensionless variables:
\bsubeqs\label{eq:dimensionless-var}
\begin{eqnarray}
\label{eq:dimensionless-var-w}
w  &\equiv& (e\,f)^2\;W = (y_0)^2+y^2\,,
\\[2mm]
\label{eq:dimensionless-var-y}
y  &\equiv& e\,f\;Y\,,
\\[2mm]
\label{eq:dimensionless-var-y0}
y_0 &\equiv& e\,f\;b\,.
\end{eqnarray}
\esubeqs

Inserting the above \textit{Ans\"{a}tze}
into the Einstein and matter field equations
from the action \eqref{eq:action-omegamu} gives the corresponding reduced
expressions. The reduced field equation (13a) from
Ref.~\cite{Klinkhamer2014-prd} contains, however, an error~\cite{Guenther2017}.
The corrected reduced field equations correspond to
the following three dimensionless ordinary differential equations (ODEs):
\bsubeqs\label{eq:final-ODEs-new-form}
\begin{eqnarray}
\label{eq:final-ODEs-sigma-corrected-new-form}
\hspace*{-5mm}
4\,w\,\sigma'(w)
&=&
+\sigma(w)\,\left[
\left[1-\sigma^2(w)\right]
+\widetilde{\eta}\, \frac{2}{w}\,
\Big(A(w)\,\sigma^2(w) + C(w)\,\left[w\,F'(w)\right]^2\Big)
\right]\,,
\\[2mm]
\label{eq:final-ODEs-mu-new-form}
\hspace*{-5mm}
4\,w\,\mu'(w)
&=&
-\mu(w)\,\left[
\left[1-\sigma^2(w)\right]
+ \widetilde{\eta}\, \frac{2}{w}\,
\Big(A(w)\,\sigma^2(w) - C(w)\,\left[w\,F'(w)\right]^2\Big)
\right]\,,
\\[2mm]
\label{eq:final-ODEs-F-new-form}
\hspace*{-5mm}
w^2\,F''(w)
&=&
+\frac{1}{C(w)}\,\sigma^2(w)\,\sin F(w)
\,\left( \sin^2\frac{F(w)}{2}+\frac{w}{2}  \right)
\nonumber\\&&
-\frac{1}{2}\,\sigma^2(w)\,w\,F'(w)
\,\left[1-4\,\widetilde{\eta}\,\frac{1}{w}\, \sin^2\frac{F(w)}{2}
 \,\left(\sin^2\frac{F(w)}{2} +w  \right) \right]
\nonumber\\&&
-\frac{1}{C(w)}\,w\,F'(w)
\,\Big[w\,F'(w)\,\sin F(w)+w \Big]
\,,
\end{eqnarray}
\esubeqs
with definitions~\cite{Guenther2017}
\bsubeqs\label{eq:A-B-defs}
\beqa
\label{eq:A-def}
A(w) &\equiv& 2\,\sin^2\frac{F(w)}{2}
\left(\sin^2\frac{F(w)}{2}+w\right)  \,,
\\[2mm]
\label{eq:B-def}
B(w) &\equiv&   w^2\, C(w)
     \equiv     w^2\, \left(4\,\sin^2\frac{F(w)}{2}+w\right) \,.
\eeqa
\esubeqs
The prime in \eqref{eq:final-ODEs-new-form}
stands for differentiation with respect to $w$.

The ODEs \eqref{eq:final-ODEs-new-form} can be solved numerically with
boundary conditions \eqref{eq:hedgehog-Ansatz-bcs}
and appropriate boundary conditions $\mu(b^2)$ and $\sigma(b^2)$.
For a given value of $\widetilde{\eta}$, the solution
space has been found to be two-dimensional~\cite{Guenther2017},
characterized by the coefficients $\{k,\,l\}$
of the asymptotic behavior
$F(w)\sim k/w$ and $\sigma(w) \sim 1+l/(2\sqrt{w})$
for $w\to\infty$.

\section{Remnant spacetime defect}\label{sec:Remnant}

As our focus will be on the asymptotic Schwarzschild mass of the
spacetime defect
(considered to be a possible left-over from an earlier phase),
we introduce the following dimensionless mass-type variable:
\beq\label{eq:l-of-w-def}
l(w) \equiv \sqrt{w}\,\left(1-\frac{1}{\sigma^{2}(w)}
\right)\,,
\eeq
which corresponds to a Schwarzschild-type behavior
of the square of the metric function,
$\sigma^{2}(w)=1/[1-l(w)/\sqrt{w}\,]$.
From the ODE \eqref{eq:final-ODEs-sigma-corrected-new-form}, we get
\beqa\label{eq:l-ODE}
l'(w)  &=&
\widetilde{\eta}\,w^{-3/2}\;
\left(A(w)+C(w)\,\frac{\left[w\,F'(w)\right]^2}{\sigma^{2}(w)} \right)\,,
\eeqa
where the prime again denotes the derivative $d/dw$
and the auxiliary functions $A(w)$ and $C(w)$ have already been
defined in \eqref{eq:A-B-defs}.
As the right-hand side of \eqref{eq:l-ODE} is nonnegative,
the interpretation is that the mass-type variable $l(w)$
can only increase by the addition of
positive energy density from the matter fields.

Still, the boundary condition on $\sigma(w)$ at $w=(y_0)^2$
[or, equivalently, on $l(w)$ at $w=(y_0)^2$]
is essentially different from what happens
with a relativistic star in a simply-connected space.
Recall that,
for an $\mathbb{R}\times \mathbb{R}^3$ spacetime manifold with
standard radial coordinate $r\in [0,\,\infty)$,
the metric at the origin $r=0$ of the relativistic star
is Minkowskian, as the local
contribution of the matter vanishes (assuming a finite
energy density at the origin).
With the standard spherically symmetric metric
over $\mathbb{R}\times \mathbb{R}^3$,
\beq\label{eq:Minkowski-metric}
ds^2 =
 -\big[\widehat{\mu}(r)\big]^2\, dt^2
 +\big[\widehat{\sigma}(r)\big]^2\,dr^2
+r^2\, \Big[d\theta^2+\sin^2 \theta\, d\phi^2 \Big]\,,
\eeq
the $rr$-component metric function $\widehat{\sigma}(r)$ then
has boundary condition
\beq
\widehat{\sigma}(0)=1\,,
\eeq
and the same holds for the $tt$-component metric function,
$\widehat{\mu}(0)=1$.
Note that $\widehat{\sigma}(0)\ne 1$ in the
metric \eqref{eq:Minkowski-metric} would give rise
to a conical singularity.

For the nonsimply-connected manifold $M_3$ of our defect,
the $YY$-component metric function $\sigma(W)$
can \emph{a priori} take any value at $W=b^2$\,:
\beq\label{eq:sigma-bc}
\sigma(b^2) \in (0,\,\infty)\,,
\eeq
where the value zero has been excluded,
in order that the field equations be well-defined
at $Y=0$ (see Sec.~3.3.1 of Ref.~\cite{Guenther2017}).
Note that, as long as $b^2 >0$,
the function value \eqref{eq:sigma-bc}
in the metric \eqref{eq:metric-Ansatz-W-definition}
does not give rise to singularities of the Ricci and
Kretschmann curvature scalars;
see Eqs.~(B1a) and (B1b) of Ref.~\cite{Klinkhamer2014-prd}
for the reduced expressions of these scalars.

We now proceed as follows.
In Sec.~\ref{sec:Critical-defect-scale}, we
describe the type of solutions obtained from
the ``standard'' boundary condition $\sigma(b^2)=1$.
In Sec.~\ref{sec:Antigravity}, we then give an argument which
demands, in certain cases, a nonstandard boundary condition
$\sigma(b^2)<1$, corresponding to a negative
mass-type variable $l(b^2)$ according to \eqref{eq:l-of-w-def}.

\section{Critical defect scale}
\label{sec:Critical-defect-scale}

\begin{figure}[p] 
\includegraphics[width=0.70\linewidth] 
{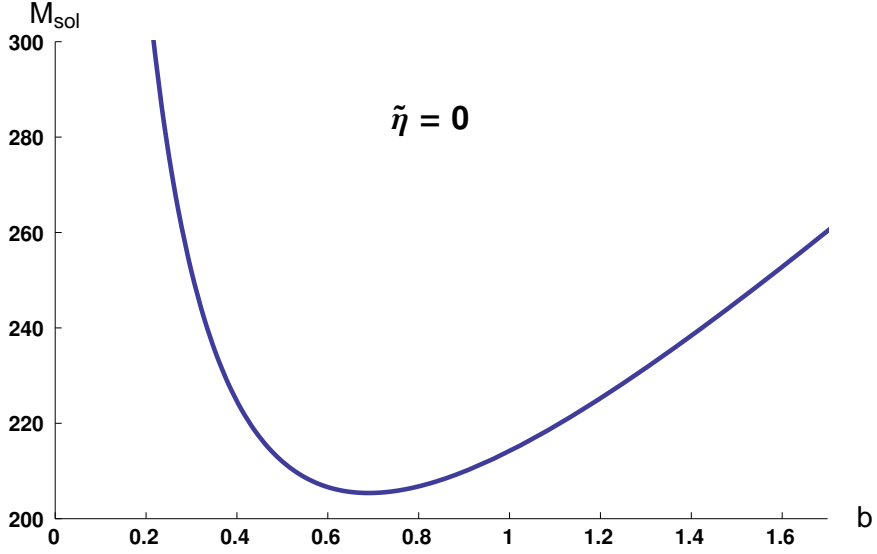}
\caption{Soliton mass $M_\text{sol}$ in units $f/e$
vs. defect scale $b$ in units $1/(ef)$,
for the case of $G_{N}=0$ and $f\ne 0$,
corresponding to a vanishing coupling constant
$\widetilde{\eta}\equiv 8\pi\, G_{N}\, f^2$.}
\label{fig01}
\end{figure}

The Skyrmion-type \textit{Ansatz} \eqref{eq:hedgehog-Ansatz},
with unit winding number, behaves as follows at the defect surface $Y=0$:
\beq\label{eq:Omega-at-Y=0}
\lim_{W\rightarrow b^2}\Omega(W)
=-\openone_3+2\,\widehat{x}\otimes \widehat{x}\,,
\eeq
which, as emphasized in Ref.~\cite{Klinkhamer2014-prd},  
is consistent with the required antipodal identification
(recall $\mathbb{R}P^2 \sim S^2/\mathbb{Z}_2$).
This behavior causes a divergence of the soliton mass
as the defect scale $b$ drops to zero  (Fig.~\ref{fig01}),
at least in the absence of gravity or $\widetilde{\eta}=0$.
The soliton mass is, in fact, defined by
\beq\label{eq:M-sol}
M_{\text{sol}}\equiv \int_{M_3} d^3x\sqrt{-g}\,T_{00}\,,
\eeq
and, for $\widetilde{\eta}=0$, behaves as follows:
\bsubeqs\beqa
M_\text{sol} &\to&\infty\,,\quad\text{for\;\;}b\rightarrow 0^{+}  \,,
\\[2mm]
M_\text{sol} &\to&\infty\,,\quad\text{for\;\;}b\rightarrow \infty  \,.
\eeqa\esubeqs
Starting from $b=0^{+}$,
the soliton mass drops to a minimum value around $b=0.7/(ef)$
and then grows again as $b$ increases (Fig.~\ref{fig01}).

If gravity is turned on ($\widetilde{\eta}\neq 0$),
the behavior of the solution is different.
The \textit{Ansatz} metric has already been given in \eqref{eq:metric-Ansatz}
and involves two functions,
$\mu(W)$ and $\sigma(W)$.
The vacuum solution for the field $\sigma(W)$ has
the Schwarzschild-type form, consistent
with Birkhoff's theorem (cf. Sec.~4 of Ref.~\cite{Klinkhamer2014-mpla}),
\beq\label{eq:Schwarzschild-type-solution}
\sigma^2(W)\;\Big|^\text{(vac)}=\frac{1}{1-R_{S}/\sqrt{W}}\,,
\eeq
where $R_{S}\equiv 2\, G_{N} \,M$ is the Schwarzschild radius.
For a nonvacuum solution, we can write
\beq\label{eq:MS-def}
\sigma^2(W)=\frac{1}{1-2\, G_{N} \,M_S(W)/\sqrt{W}}\,.
\eeq
The function $M_S(W)$ can essentially be interpreted as 
the mass contained within a sphere of squared radius $W$
and will be discussed further below. 
Note that the dimensionless version of $2\, G_{N} \,M_S(W)$
corresponds to our previous variable $l(w)$ from \eqref{eq:l-of-w-def},
\beq\label{eq:l-w-and-MS-W}
l(w) = 2\, G_{N} \,M_S(W)\,e\,f\,,
\eeq
where $w$ and $W$ are related by \eqref{eq:dimensionless-var-w}.

The Arnowitt--Deser--Misner (ADM)
mass~\cite{ADM1959} is now obtained by the limit
\beq\label{eq:M-ADM}
M_{\text{ADM}} =\lim_{W\rightarrow\infty}M_S(W)
\eeq
and equals the Komar mass~\cite{Komar1963}  for the case considered
(see Sec.~2.3.1.2 of Ref.~\cite{Guenther2017}).

\begin{figure}[p] 
\includegraphics[width=0.70\linewidth]%
{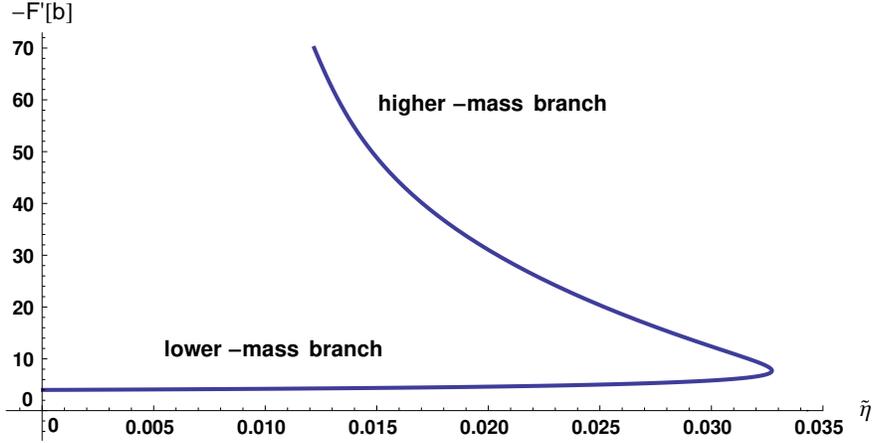}
\caption{Two branches of the Skyrmion spacetime defect solution:
matter-function slope $F'(b^2)$  vs. coupling constant $\widetilde{\eta}$,
for defect scale $b = 0.34/(ef)$ and metric-function
boundary condition $\sigma(b^2)=1$. At this value of $b$, the critical value of the coupling constant, $\widetilde{\eta}_\text{crit}$,
is approximately equal to $0.033$.}
\label{fig02}
\end{figure}

The model from Sec.~\ref{sec:Theory-etc},
with the standard boundary condition $\sigma(b^2)=1$,
gives rise to two branches of solutions  (Fig.~\ref{fig02})
that coalesce at a point
characterized by a critical defect scale $b_\text{crit}$
and a critical coupling $\widetilde{\eta}_\text{crit}$.
(The gravitating $SU(2)$ Skyrmion also has two branches;
cf. Fig.~1  and Table~1 of Ref.~\cite{BizonChmaj1992}.)
The branch with larger values of $\vert F'(b^2)\vert$
in Fig.~\ref{fig02}
has larger ADM masses and presumably corresponds to unstable solutions.
We will restrict our analysis to the branch with smaller values of
$\vert F'(b^2)\vert$ and smaller ADM masses.
(The last two sentences correct some erroneous statements
in Ref.~\cite{Klinkhamer2014-prd}, at the bottom of the left column
\mbox{on p. 5.)}

\begin{figure}[p] 
\includegraphics[width=0.70\linewidth]
{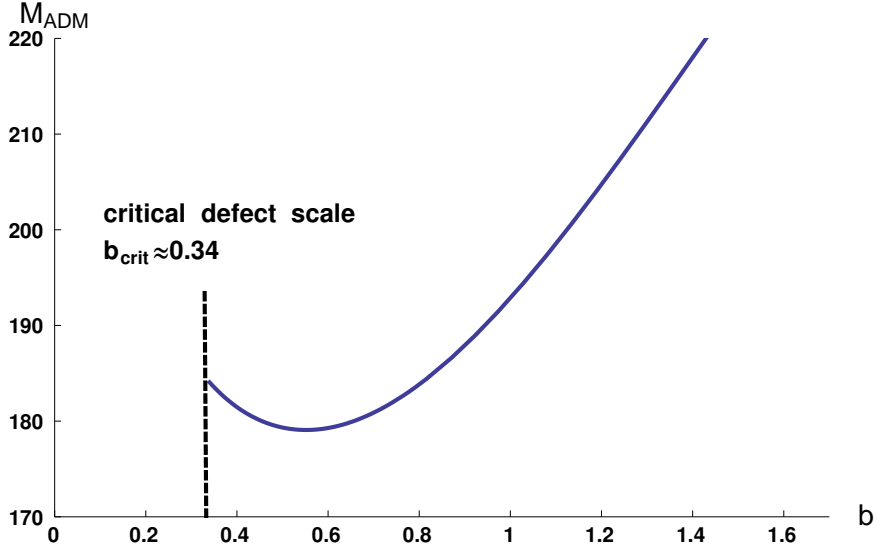}
\caption{Gravitational mass $M_\text{ADM}$
in units $f/e$ vs. defect scale $b$ in units $1/(ef)$,
for coupling constant $\widetilde{\eta}=0.033$
and boundary condition $\sigma(b^2)=1$.
These results correspond to the lower-mass branch of Fig.~\ref{fig02}.}
\label{fig03}
\end{figure}

\begin{figure}[p] 
\includegraphics[width=0.70\linewidth]
{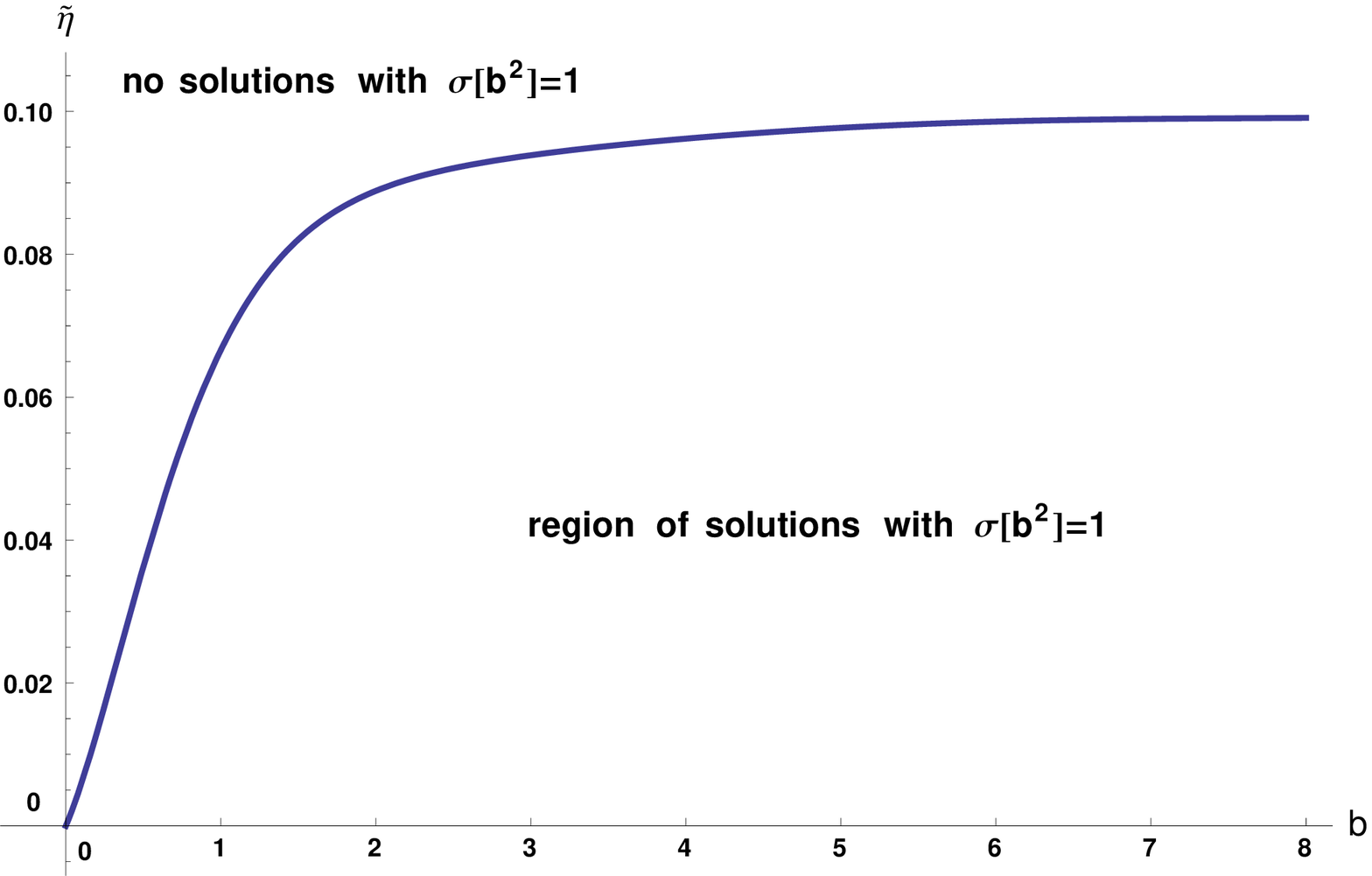}
\caption{Curve of the critical defect scale $b_\text{crit}$
[in units of $1/(e f)$] and corresponding critical coupling constant
$\widetilde{\eta}_\text{crit}$, where the boundary condition is
fixed to $\sigma(b^2)=1$. Above the curve, there are
no globally regular solutions with $\sigma(b^2)=1$.}
\label{fig04}
\end{figure}

From now on, we will refer to the coordinate-dependent quantity $M_S(W)$
from \eqref{eq:MS-def} as the  ``effective mass.''
Therefore, the boundary condition $\sigma(b^2)=1$ can be understood
as having a vanishing effective mass at the defect surface.
In this case, one expects that, for a
given coupling constant $\widetilde{\eta}\neq 0$, the solution ceases
to exist below some critical value of the defect scale.
The numerical solution of the ODEs indeed shows this behavior
(Fig.~\ref{fig03}):
starting from $b$ values near the minimum ADM mass,
the soliton energy increases as $b$ decreases,
but, at $b=b_{\text{crit}}$,
the solution collapses, possibly even before an event horizon is formed.
(The collapse of the gravitating $SU(2)$ Skyrmion
has been discussed in, e.g., Ref.~\cite{BizonChmaj1992}.)
Note that, in the absence of gravity ($\widetilde{\eta}=0$),
the collapse of the unit-winding-number
$SO(3)$ Skyrmion is impossible and the soliton energy increases
to infinity for $b\to 0^{+}$, as shown by Fig.~\ref{fig01}.

With gravity present, the order of magnitude of $b_{\text{crit}}$
can be obtained from the following condition:
\beq\label{eq:b-crit-cond}
b =
\kappa\; 2\,G_{N}\,M_{\text{sol},\,\widetilde{\eta}=0}(b^2)/c^2
+\text{O}(\widetilde{\eta}^2)\,,
\eeq
where we have temporarily reinstated the light velocity $c$.
In \eqref{eq:b-crit-cond}, $\kappa$ is a positive
constant of order unity and $M_{\text{sol},\,\widetilde{\eta}=0}$
is given by the integral \eqref{eq:M-sol}
for the case $\widetilde{\eta}=0$.
The numerical results from Figs.~\ref{fig01} and \ref{fig03}
are consistent with \eqref{eq:b-crit-cond}
for $\kappa\approx 27/50$\,:
the first term on the right-hand side
of \eqref{eq:b-crit-cond} gives approximately $0.34/(ef)$
for $\widetilde{\eta}=0.033$
and $M_{\text{sol},\,\widetilde{\eta}=0}\approx 240\,f/e$
from the curve of Fig.~\ref{fig01} at $b=0.34/(ef)$,
while the left-hand side of \eqref{eq:b-crit-cond}
corresponds to the direct numerical result
$b_{\text{crit}} \approx 0.34/(ef)$ from Fig.~\ref{fig03}.
In Fig.~\ref{fig04}, we show the numerical results for the
dependence of $b_{\text{crit}}$
on the coupling $\widetilde{\eta}_{\text{crit}}$
at fixed boundary condition $\sigma(b^2)=1$.
Above the curve of Fig.~\ref{fig04}, there are no globally regular 
solutions with $\sigma(b^2)=1$ or $M_{S}(b^2)=0$.

\section{Antigravity}\label{sec:Antigravity}

Now, assume that the spacetime defects
are remnants from some quantum-spacetime (QST) phase,
as discussed in Sec.~\ref{sec:Introduction}.
For simplicity, also assume
that this phase determines a single characteristic scale
of the defects, $b_\text{QST}>0$.
There are two possibilities to consider.
The first possibility is that the characteristic scale,
for a fixed value of $\widetilde{\eta}$,
corresponds to $b_\text{QST}^\text{(1)}\geq b_{\text{crit}}$
[lying, for example, in the $b>0.34/(ef)$ region of Fig.~\ref{fig03}\,].
With the boundary condition
$\sigma\big(b^2\big)=1$ for $b=b_\text{QST}^\text{(1)}$,
the ADM mass is then always positive:
the effective-mass function $M_S(W)$ is zero at
$W=\big(b_\text{QST}^\text{(1)}\big)^2$
\mbox{and increases monotonically with $W$.}

The second possibility is that the characteristic scale,
for a fixed value of $\widetilde{\eta}$, corresponds to
\beq\label{eq:bQST-small}
b_\text{QST}^\text{(2)}< b_{\text{crit}}\,,
\eeq
where, from now on, the suffix ``(2)'' on the left-hand-side
expression will be dropped
and $b_{\text{crit}}$ on the right-hand side
has been given explicitly in Fig.~\ref{fig04}
and implicitly by \eqref{eq:b-crit-cond}.
Then, in order to have globally regular solutions and to
avoid gravitational collapse, we must have
\beq\label{eq:sigma-defect-small}
0<\sigma(b^2)\;\Big|_{b=b_\text{QST}}<1\,.
\eeq
The condition \eqref{eq:sigma-defect-small}
translates into $M_{S}(b_\text{QST}^2)<0$,
so that the effective mass at the defect surface is negative.
Adding a sufficiently negative effective mass to the defect decreases
the critical scale of the defect significantly,
allowing for
$b_\text{QST}  \geq b_{\text{crit}}^\text{(new)}$.
The effective-mass function $M_S(W)$ is still monotonic but
starts from a negative value at $W=b^2$.

\begin{figure*}[p]
\vspace*{-4mm}
\includegraphics[width=0.8\textwidth]
{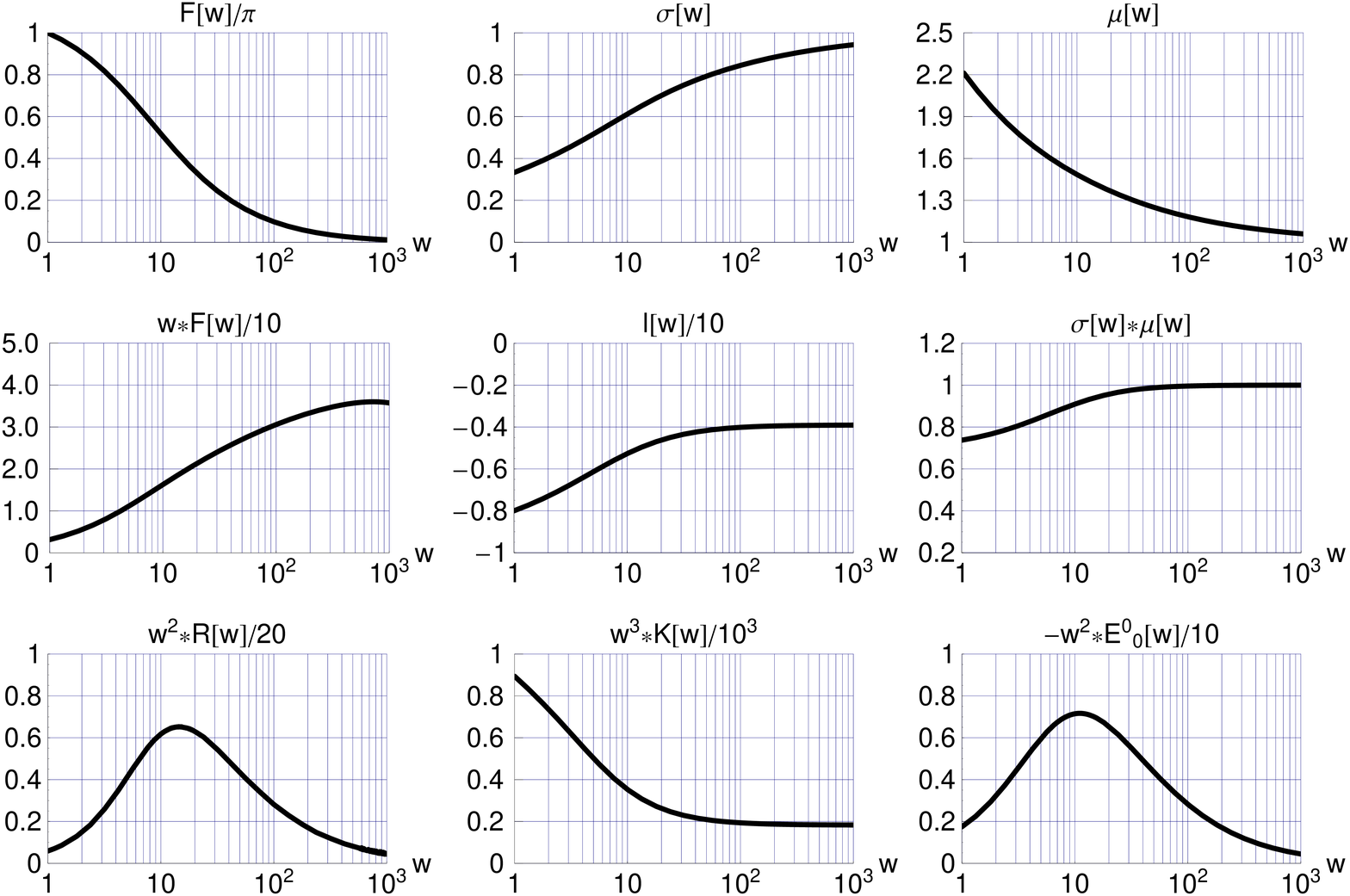}
\vspace*{-4mm}
\caption{
(Top row)
\textit{Ansatz} functions $F(w)$, $\sigma(w)$, and $\mu(w)$
of the numerical solution
of the reduced field equations \eqref{eq:final-ODEs-new-form}.
The parameters are $\widetilde{\eta}\equiv 8\pi\, G_{N}\, f^2  =1/10$ and
$y_0 \equiv ef b =1$.
The boundary conditions at the defect surface $w=(y_0)^2=1$ are:
$F=\pi$,
$F^\prime=-0.323978148$,
$\sigma=1/3$,
and  $\mu=2.21176$.
(Middle row)
Derived functions $w\,F(w)$,
$l(w)\equiv \sqrt{w}\,\left[1-1/\sigma^{2}(w)\right]$,
and $\sigma(w)\,\mu(w)$.
(Bottom row) Dimensionless Ricci curvature scalar $R(w)$,
dimensionless Kretschmann curvature  scalar $K(w)$,
and negative of the
00 component of the dimensionless Einstein tensor $E^{\mu}_{\;\;\nu}(w)$
$\equiv$ $R^{\mu}_{\;\;\nu}(w) -(1/2)\,R(w)\,\delta^{\mu}_{\;\;\nu}$,
where the explicit expressions are given in
App.~B of Ref.~\cite{Klinkhamer2014-prd}.
}
\label{fig5}  
\end{figure*}

\begin{figure*}[p]
\vspace*{-4mm}
\includegraphics[width=0.8\textwidth]
{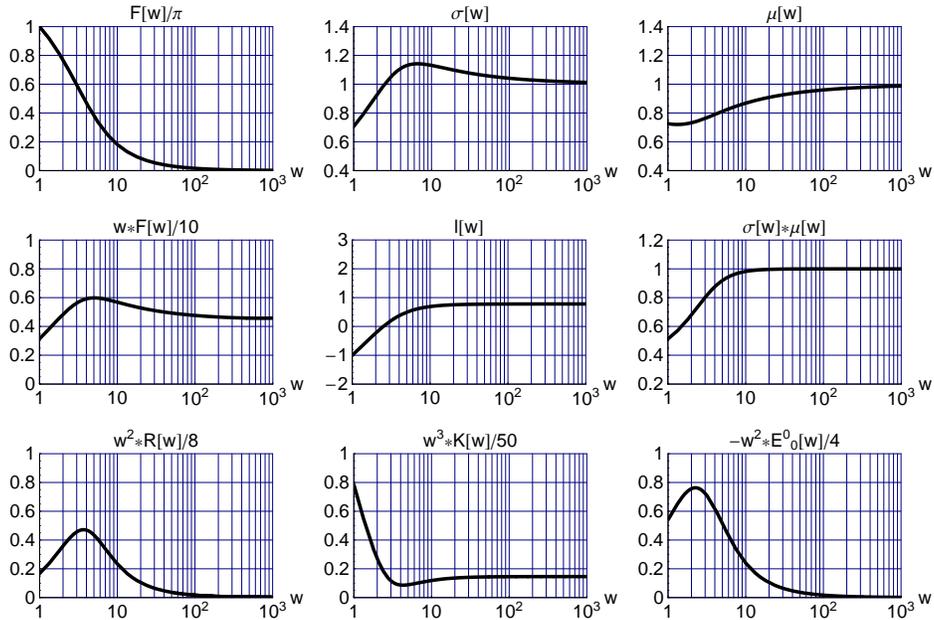}
\vspace*{-4mm}
\caption{Same as Fig.~\ref{fig5}, but now with the following
boundary conditions at the defect surface $w=(y_0)^2=1$:
$F=\pi$,
$F^\prime=-0.82561881304$,
$\sigma=1/\sqrt{2}$,
and  $\mu=0.725818$.
}
\label{fig6} 
\end{figure*}

For finite values of $\widetilde{\eta}$,
the ODEs \eqref{eq:final-ODEs-new-form}
with nonstandard boundary condition $\sigma(b_\text{QST}^2)<1$
from \eqref{eq:sigma-defect-small} [or $M_{S}(b_\text{QST}^2)<0$]
can be solved numerically and give an
ADM mass \eqref{eq:M-ADM} which can be negative, zero, or positive.
Two  numerical solutions are presented in
Figs.~\ref{fig5} and \ref{fig6} for parameters
$\{\widetilde{\eta},\,b\}=\{1/10,\,1/(ef)\}$,
which correspond to a point above the curve of Fig.~\ref{fig04}.
These numerical solutions have,
depending on the value of the boundary condition
$\sigma(b^2)<1$, negative and positive ADM mass, as shown by
the respective $l(w)$ panels in Figs.~\ref{fig5} and \ref{fig6}.
The $\sigma(b^2)$ dependence of the two-dimensional solution
space, at three fixed values of $\widetilde{\eta}$,
is given in Fig.~4.4 of Ref.~\cite{Guenther2017}.

\renewcommand{\baselinestretch}{1.22}
\renewcommand{\baselinestretch}{1.26}

Next, consider
the special case $\widetilde{\eta}=0^{+}$, or explicitly
\beq
f^2 \ll      \big(E_\text{planck}\big)^2
    \equiv   1/(8\pi G_{N})  \,,
\eeq
and assume a small enough value of the remnant-defect
scale \eqref{eq:bQST-small}.
From the definition \eqref{eq:l-w-and-MS-W} and the ODE \eqref{eq:l-ODE},
there is then the following behavior: the value of $M_S(W)$ stays
close to the negative value of $M_{S}(b_\text{QST}^2)$
from \eqref{eq:sigma-defect-small}
and the asymptotic ADM mass \eqref{eq:M-ADM} is negative.
For this special case, the negative ADM mass has been established
by analytically solving ODE \eqref{eq:l-ODE} for $\widetilde{\eta}=0$,
with the trivial solution $l(w)=\text{const}=l(y_0^2)$.
But the $F(w)$ profile still requires the numerical solution of
ODE \eqref{eq:final-ODEs-F-new-form}
with $\sigma^2(w)$ replaced by
$\widetilde{\sigma}^2(w) \equiv 1/[1-l/\sqrt{w}\,]$ 
for constant $l \equiv l(y_0^2) <0$,
where $\widetilde{\sigma}(w)$ and
$\widetilde{\mu}(w)=1/\widetilde{\sigma}(w)$ solve
the ODEs \eqref{eq:final-ODEs-sigma-corrected-new-form}
and \eqref{eq:final-ODEs-mu-new-form} at $\widetilde{\eta}=0$.
An explicit choice for $b_\text{QST}$ will be discussed in
App.~\ref{app:Planck-scale-defects},
which also contains numerical results for the $F(w)$ profile.
Additional numerical results are given in
App.~\ref{app:Additional-numerical-results}.

Generally speaking, a negative ADM mass of the Skyrmion spacetime defect
corresponds to a negative active gravitational mass
(see, e.g., Ref.~\cite{Bondi1957} for
a discussion of negative masses in general relativity).
The globally regular soliton then repels a distant test particle.
Hence, there is ``antigravity'' from this particular
Skyrmion spacetime defect,
as long as the defect scale $b_\text{QST}$ and the
coupling constant $\widetilde{\eta}$ are small enough.

\section{Discussion}\label{sec:Discussion}

In this article, we have explained why
a particular Skyrmion spacetime defect solution
requires a negative ADM mass for its existence.
In short, a nonstandard boundary
condition on one of the metric functions is needed,
$\sigma(b^2)<1$,  in order to avoid gravitational
collapse and, for a small enough value of the
coupling constant $\widetilde{\eta}\equiv 8\pi\, G_{N}\, f^2$,
this boundary condition at the defect surface directly gives
a negative ADM mass at spatial infinity.
Whether or not such negative-gravitational-mass defects are present
in the actual Universe depends on the nature of
the quantum-spacetime phase
and its supposed transition to an emerging classical spacetime.
Needless to say, this quantum-spacetime phase is \emph{terra incognita}.

Returning to the spacetime defect classical solution by itself,
we have six general remarks on the physics of the
negative-gravitational-mass soliton.
First, the negative ADM mass found in Sec.~\ref{sec:Antigravity}
is not due to ponderable matter but to the gravitational
fields at the defect surface $W=b^2$ (or $Y=0$).
Indeed, the vacuum solution \eqref{eq:Schwarzschild-type-solution}
with $R_S \equiv 2 G_N M$ does not have a point
where the curvature diverges and where ponderable matter can be
thought to reside (the numerical value of constant $M$
is set by the gravitational fields themselves
rather than by the ponderable matter).
This behavior differs from that of
the Schwarzschild solution
[given by metric \eqref{eq:Minkowski-metric}
with functions $\widehat{\mu}(r)=1/\widehat{\sigma}(r)=
(1-2 G_N \widehat{M}/r)$], which
does have a point ($r=0$)
where the curvature diverges and where ponderable matter can be
thought to reside
(this ponderable matter gives a positive numerical value for the
constant $\widehat{M}$).

Second,
the Skyrmion-spacetime-defect metric
from \eqref{eq:metric-Ansatz-W-definition}
is degenerate: $\det g_{\mu\nu}=0$ at the defect surface $Y = 0$,
which corresponds to an $\mathbb{R}P^2$ submanifold. 
This degenerate metric
makes that the singularity~\cite{Gannon1975}
and positive-mass~\cite{SchoenYau1979} theorems
are not directly applicable, allowing for a
negative gravitational mass of the
Skyrmion spacetime defect solution; see Sec.~3.1.5
of Ref.~\cite{Guenther2017} for further discussion.
The special feature of the Skyrmion spacetime defect solution
is that certain geodesics at the defect surface $Y = 0$
cannot be continued uniquely.

Third, the heuristic understanding of why there is degeneracy 
($\det g_{\mu\nu}=0$) at the defect surface $Y = 0$ is as follows.
The two-dimensional defect surface at $Y=0$ corresponds
to the projective plane $\mathbb{R}P^2$.
Now, it is a well-known result that $\mathbb{R}P^2$ cannot be
differentiably embedded in $\mathbb{R}^3$ (i.e., without intersections); 
cf. p.~40 in Ref.~\cite{Nakahara1990}
and Fig.~3 in Ref.~\cite{Klinkhamer2014-mpla}. 
For us, this result then suggests that the third extra dimension $Y$ 
emerging from the two-dimensional defect surface
(local coordinates $X$ and $Z$)
must have a vanishing metric component
$g_{YY}$ at $Y=0$, which is precisely the structure
of our metric \textit{Ansatz} \eqref{eq:metric-Ansatz-W-definition}
for a finite value of $\sigma(b^2)$.  
If the metric component $g_{YY}$ at $Y=0$ does not vanish, then there
occur curvature singularities, as discussed in Sec.~5.2 of
Ref.~\cite{Schwarz2010}.
Note that our defect-surface condition $\det g_{\mu\nu}=0$ 
is invariant under nonsingular coordinate transformations.

Fourth,
having established that the
Skyrmion spacetime defect has a negative active gravitational mass,
as long as the defect scale $b$ and the coupling constant $\widetilde{\eta}$
are small enough,
the question arises as to the value of its inertial mass.
A positive inertial mass of the soliton-type solution would
correspond to a dynamic violation of the equivalence principle,
but this would not be altogether surprising as the solution
is known to violate the standard elementary flatness property
at the defect surface $Y=0$; see the paragraph under Eq.~(2.29)
in Ref.~\cite{KlinkhamerSorba2014} for details.

Fifth, 
the negative-gravitational-mass Skyrmion spacetime defect
may have other unusual properties, such as the nonstandard
parity eigenvalues of scalar-field scattering solutions
(a summary of the results appears in Sec.~IV of Ref.~\cite{KlinkhamerSorba2014}).

Sixth, and finally,
it remains to perform a rigorous stability analysis of the
negative-gravitational-mass Skyrmion spacetime defect solution.

As to possible applications of antigravity by spacetime defects,
two areas come to mind: cosmology (dark matter and dark energy)
and condensed matter physics (analogue-gravity systems).

\vspace*{-0mm}
\section*{\hspace*{-5mm}Acknowledgments}
\vspace*{-0mm}\noindent
It is a pleasure to thank M. Guenther and M. Kopp 
for useful comments on the manuscript.

\begin{appendix}
\section{Planck-scale defects}
\label{app:Planck-scale-defects}
\vspace*{-0mm}

In Sec.~\ref{sec:Antigravity}, we have given a general discussion of
the antigravity effect coming from a small enough
defect scale $b$, but we did not specify the actual size of the
defect scale. In this appendix, we discuss a particular
choice for the characteristic scale $b_\text{QST}$
of the defects, where the defects are considered to be
remnants from an earlier quantum-spacetime (QST) phase.

First, restrict the matter sector of \eqref{eq:action-omegamu}
to the following parameter domain:
\bsubeqs\label{eq:matter-theory-f2bound-ebound}
\beqa\label{eq:matter-theory-f2bound}
f^2  &\ll&    \big(E_\text{planck}\big)^2
     \equiv   1/(8\pi G_{N})
     \approx  \big(2.44\times 10^{18}\;\text{GeV}\big)^2\,,
\\[2mm]
\label{eq:matter-theory-ebound}
e    &\leq& 1\,,
\eeqa
\esubeqs
where the first inequality corresponds to $\widetilde{\eta}\ll 1$
and where the origin of the upper bound on the coupling constant $e$ will
be explained shortly. Next, assume that the
characteristic scale of the defects is given by the Planck length,
\beq\label{eq:bQST-planck}
b_\text{QST} = l_\text{planck}
             \equiv \sqrt{8\pi G_{N}\, \hbar/c^3}
             \approx 8.10 \times 10^{-35}\;\text{m}\,,
\eeq
which may very well be a typical scale of the quantum-spacetime phase
(note the appearance of $\hbar$ in the square root of the above
expression).

\begin{figure*}[t]
\vspace*{-4mm}
\includegraphics[width=0.8\textwidth]
{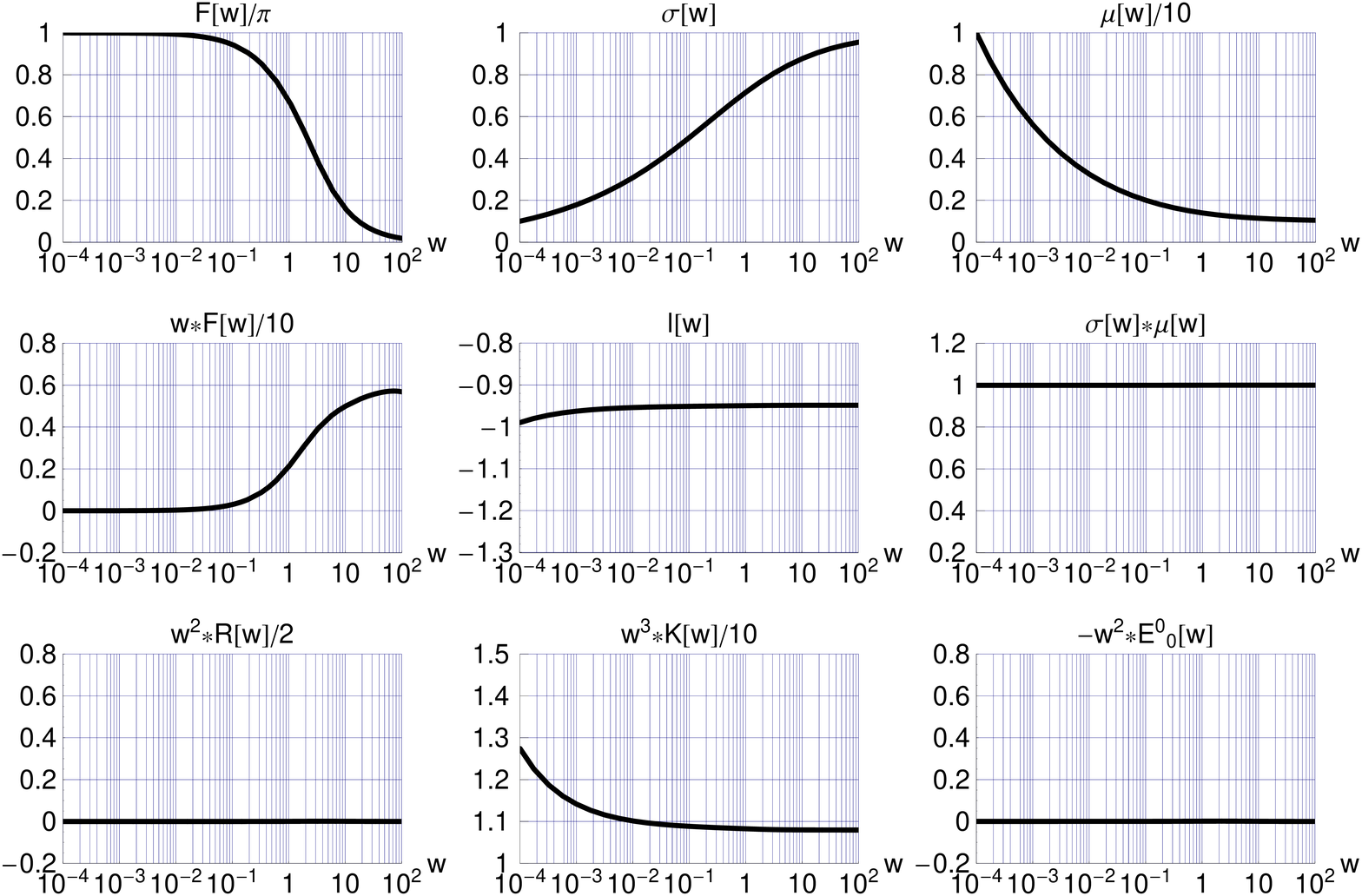}
\vspace*{-0.00006mm}
\caption{(Top row)
\textit{Ansatz} functions $F(w)$, $\sigma(w)$, and $\mu(w)$
of the numerical solution
of the reduced field equations \eqref{eq:final-ODEs-new-form}.
The parameters are
$\widetilde{\eta}\equiv 8\pi\, G_{N}\, f^2=10^{-4}$, $e=1$, and
$y_0 \equiv ef b = ef l_\text{planck}=  10^{-2}$.
The boundary conditions at the defect surface $w=(y_0)^2=10^{-4}$ are:
$F=\pi$,
$F^\prime=-2.27604760$,
$\sigma=1/10$,
and  $\mu=9.99298$.
(Middle row)
Derived functions $w\,F(w)$,
$l(w)\equiv \sqrt{w}\,\left[1-1/\sigma^{2}(w)\right]$,
and $\sigma(w)\,\mu(w)$.
(Bottom row) Dimensionless Ricci curvature scalar $R(w)$,
dimensionless Kretschmann curvature  scalar $K(w)$,
and negative of the
00 component of the dimensionless Einstein tensor $E^{\mu}_{\;\;\nu}(w)$
$\equiv$ $R^{\mu}_{\;\;\nu}(w) -(1/2)\,R(w)\,\delta^{\mu}_{\;\;\nu}$.
}
\label{fig7}
\vspace*{-0.00002mm}
\end{figure*}

With the assumptions \eqref{eq:matter-theory-f2bound-ebound}
and \eqref{eq:bQST-planck},
condition \eqref{eq:bQST-small} holds directly
[in the terminology of the first two paragraphs
of Sec.~\ref{sec:Antigravity}, there is only the second possibility].
As explained in Sec.~\ref{sec:Antigravity},
condition \eqref{eq:bQST-small}
implies that the globally regular solution must have
$\sigma(b_\text{QST}^2)<1$ and a corresponding
negative ADM mass as given by \eqref{eq:M-ADM}.
The reason is that, with $\widetilde{\eta}\ll 1$,
the effective mass $M_S(W)$ is approximately
constant for all $W$ values between $b^2$ and infinity.
This behavior of nearly constant $M_S(W)$,
or equivalently nearly constant $l(w)$
from \eqref{eq:l-w-and-MS-W},
is already visible in Fig.~\ref{fig7} with numerical results for
$\widetilde{\eta}=10^{-4}$, $e=1$, and defect scale $b=l_\text{planck}$
from \eqref{eq:bQST-planck}.
Observe also that the behavior
of the bottom-row panels in
Fig.~\ref{fig7} rapidly approaches the behavior of the
defect vacuum solution (cf. Sec.~3 in Ref.~\cite{Klinkhamer2014-mpla})
with vanishing Ricci curvature tensor, $R_{\mu\nu}^\text{(vac)}(w)=0$,
and nonvanishing Kretschmann curvature scalar,
$K^\text{(vac)}(w)\propto M^2/w^3$.

Let us now give the promised derivation of \eqref{eq:matter-theory-ebound}.
From Fig.~\ref{fig01} for $b \to 0^{+}$, we
obtain $M_\text{sol} \propto 1/b$ or explicitly
$M_\text{sol} \sim (70/y_0)\,f/e$.
With the latter result inserted in condition
\eqref{eq:b-crit-cond} for $b_\text{crit}$,
condition \eqref{eq:bQST-small} with
$b=b_\text{QST} = l_\text{planck}$ translates
into an upper bound on the matter coupling,
\mbox{$e < \sqrt{\kappa\,70/(4\pi)} \approx 1.73$}
for $\kappa \approx 0.54$.
In the bound \eqref{eq:matter-theory-ebound}, we have just taken
the number $1$ instead of the more accurate number $1.73$.

With the actual parameter value as stated
in \eqref{eq:matter-theory-ebound},
the difference between the
defect scale $b$ from \eqref{eq:bQST-planck} and the critical
scale $b_\text{crit}$
is a finite negative number, $b-b_\text{crit}<0$,  
and, therefore, requires a
finite negative value for $\sigma(b^2)-1$,
in order to get a regular solution.
In turn, a finite negative value of $\sigma(b^2)-1$
corresponds to a finite negative value of $l(b^2)$
as defined by \eqref{eq:l-of-w-def}, which implies,
for $\widetilde{\eta}$ sufficiently close to $0$,
a finite negative value of $\lim_{w\to\infty}l(w)$,
i.e., a negative ADM mass, according to
\eqref{eq:l-w-and-MS-W} and \eqref{eq:M-ADM}.

To summarize, for the matter theory
\eqref{eq:action-omegamu} with restricted
parameters \eqref{eq:matter-theory-f2bound-ebound},
remnant spacetime defects with a length scale given by
the Planck length \eqref{eq:bQST-planck}
have a negative ADM mass, as long as they correspond to
globally regular solutions. The implicit assumption, here,
is that general relativity gives a more or less reliable
description even for extremely small length scales
as given by \eqref{eq:bQST-planck}.
But it is also possible that
the defect scale is significantly larger than the Planck length,
$b_\text{QST} = \zeta\, l_\text{planck}$
with a numerical factor $\zeta\gg 1$,
provided the matter coupling constant is correspondingly reduced,
$e \leq 1/\zeta \ll 1$.
\vspace*{-0.00004mm}

\section{Additional numerical results}
\label{app:Additional-numerical-results}
\vspace*{-0.00001mm}

The numerical results of Ref.~\cite{Klinkhamer2014-prd}
are incorrect, because one of the three ODEs used
contained some erroneous terms.
As mentioned in Sec.~\ref{sec:Theory-etc},
the ODE (13a) in Ref.~\cite{Klinkhamer2014-prd} needs to be
replaced by \eqref{eq:final-ODEs-sigma-corrected-new-form}.
In this appendix, we present the corrected numerical results.

\begin{figure*}[t]
\vspace*{-4mm}
\includegraphics[width=0.8\textwidth]
{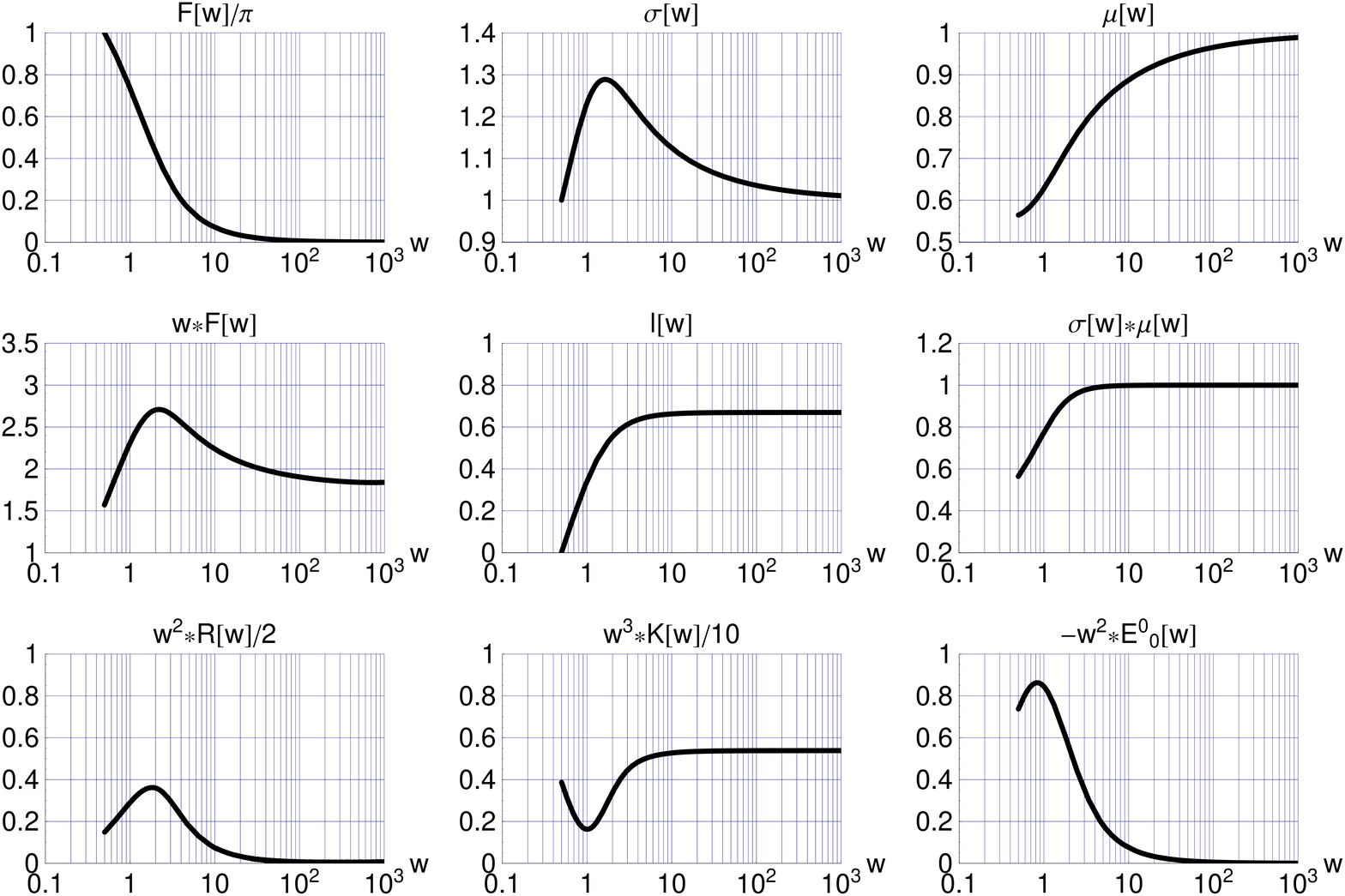}
\vspace*{-0mm}
\caption{
(Top row)
\textit{Ansatz} functions $F(w)$, $\sigma(w)$, and $\mu(w)$
of the numerical solution
of the reduced field equations \eqref{eq:final-ODEs-new-form}.
The parameters are $\widetilde{\eta}=1/20$ and $y_0=1/\sqrt{2}$.
The boundary conditions at the defect surface $w=(y_0)^2=1/2$ are:
$F=\pi$,
$F^\prime=-1.9718377138$,
$\sigma=1$,
and  $\mu=0.564337$.
(Middle row)
Derived functions $w\,F(w)$,
$l(w)\equiv \sqrt{w}\,\left[1-1/\sigma^{2}(w)\right]$,
and $\sigma(w)\,\mu(w)$.
(Bottom row) Dimensionless Ricci curvature scalar $R(w)$,
dimensionless Kretschmann curvature  scalar $K(w)$,
and negative of the
00 component of the dimensionless Einstein tensor $E^{\mu}_{\;\;\nu}(w)$
$\equiv$ $R^{\mu}_{\;\;\nu}(w) -(1/2)\,R(w)\,\delta^{\mu}_{\;\;\nu}$.
}
\label{fig8} 
\vspace*{10mm}
\vspace*{-0.00004mm}
\includegraphics[width=0.8\textwidth]
{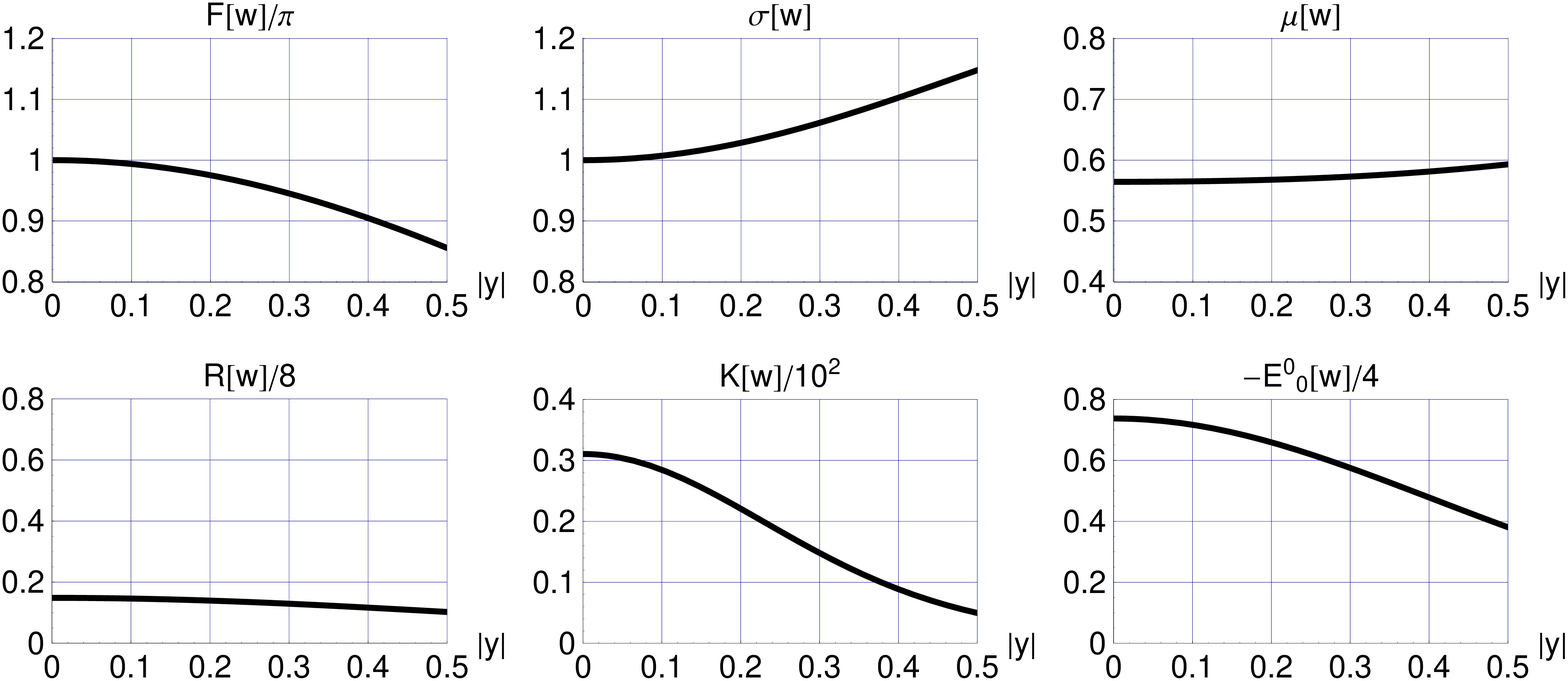}
\vspace*{-0mm}
\caption{Core behavior of the functions from Fig.~\ref{fig8},
in terms of the dimensionless quasiradial coordinate $y \in \mathbb{R}$.
The coordinate $w$ is given by $w\equiv (y_0)^2+y^2 = 1/2+y^2$.}
\label{fig9} 
\end{figure*}

Taking the parameter values
$\{\widetilde{\eta},\, b\}=\{1/20,\,(1/\sqrt{2})/(e f) \}$,
Fig.~\ref{fig8} replaces Figs.~2 and 3 of Ref.~\cite{Klinkhamer2014-prd}
and Fig.~\ref{fig9} replaces Fig.~4 of Ref.~\cite{Klinkhamer2014-prd}.
The boundary value $\sigma(b^2)=1$ in Fig.~\ref{fig8}
has been chosen to give approximately the
same asymptotic $l$ value as in Fig.~3 of
Ref.~\cite{Klinkhamer2014-prd}.
Remark that the parameter values $\{\widetilde{\eta},\, b\}$
of Fig.~\ref{fig8}
correspond to a point just below the curve of Fig.~\ref{fig04}.

Figure~\ref{fig9} shows that the \textit{Ansatz} functions of the
numerical solution are perfectly smooth at the defect surface
$Y=0$, as are the Ricci and Kretschmann curvature scalars
and the energy density.

\end{appendix}

\end{document}